# Commissioning of the new high-intensity ultracold neutron source at the Paul Scherrer Institut

**B Lauss**[4a]
**on behalf of the PSI UCN Project Team**[5]

[a]Paul Scherrer Institut, CH-5232 Villigen-PSI, Switzerland

E-mail: bernhard.lauss@psi.ch

**Abstract.** Commissioning of the new high-intensity ultracold neutron (UCN) source at the Paul Scherrer Institut (PSI) has started in 2009. The design goal of this new generation high intensity UCN source is to surpass by a factor of ~100 the current ultracold neutron densities available for fundamental physics research, with the greatest thrust coming from the search for a neutron electric dipole moment. The PSI UCN source is based on neutron production via proton induced lead spallation, followed by neutron thermalization in heavy water and neutron cooling in a solid deuterium crystal to cold and ultracold energies. A successful beam test with up to 2 mA proton beam on the spallation target was conducted recently. Most source components are installed, others being finally mounted. The installation is on the track for the first cool-down and UCN production in 2010.

## 1. Introduction

The study of fundamental properties of the neutron has strongly benefitted from the possibility to store neutrons for times of the order of the neutron β-decay lifetime. Neutron storage can be achieved via decreasing the neutron energy below the material optical potential (Fermi potential) of suitable 'storage' materials, i.e. below ~300 neV. Such ultralow kinetic energies correspond to neutron velocities below 8m/s or temperatures below 3mK; hence neutrons at such energies are termed ultracold neutrons (UCN) [1]. Typical materials used in UCN applications have high material optical potentials of up to 190 neV (stainless steel), 250 neV (Ni, Be, DLC) or 335 neV ($^{58}$Ni). On the same scale is the UCN energy gain/loss in the gravitational potential, namely 102 neV per meter height, and in a magnetic field (64 neV per Tesla). Hence UCN can be manipulated via these interactions.

The sensitive search for a permanent electric dipole moment of the neutron [2,3], sets nowadays new demands of increased UCN densities and intensities available to experiments. Consequently, new UCN sources based on superthermal UCN production are being presently considered, designed and built worldwide (see Ref.[4] for a current status).

---

[4] To whom any correspondence should be addressed.
[5] The members of the PSI UCN Project Team are listed at http://ucn.web.psi.ch .

## 2. The PSI ultracold neutron source

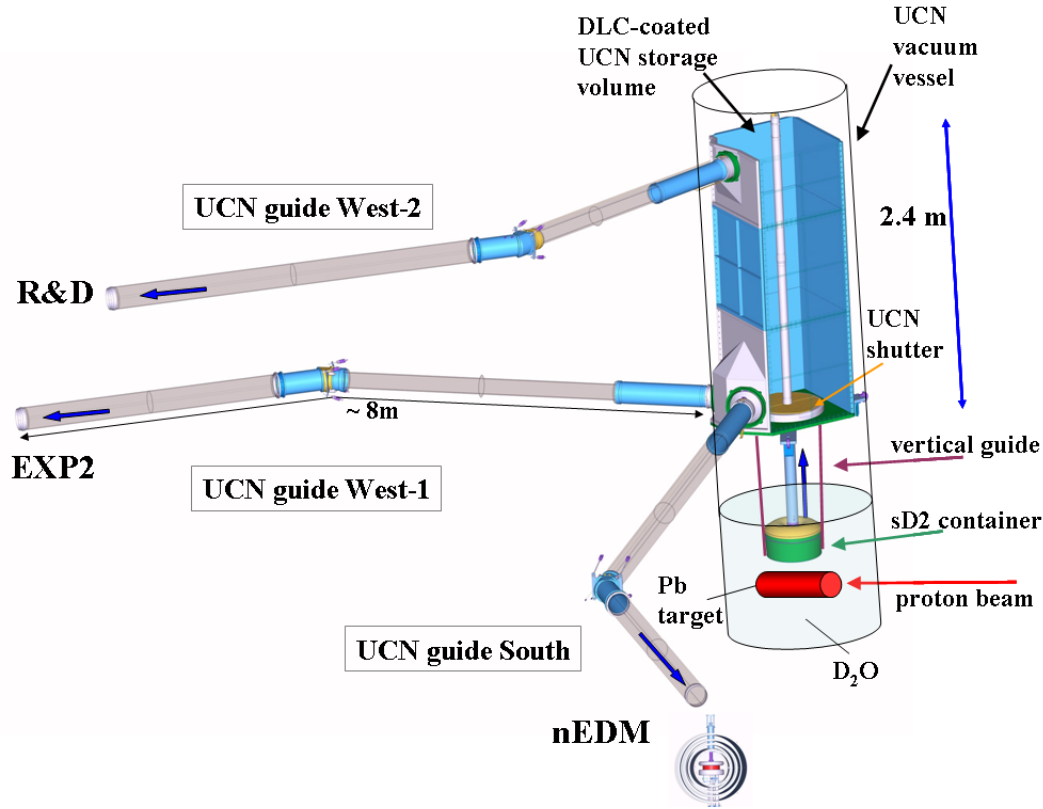

Fig.1: Sketch of the main components relevant to UCN production and transport. All the shown components are installed inside a large vacuum vessel (Fig.5a).

A spallation source for ultracold neutrons has been constructed at the Paul Scherrer Institut (PSI), Villigen, Switzerland. Commissioning has started in fall 2009. The main components of the source are sketched in Fig.1. Several stages are necessary to first produce neutrons and then slow down these neutrons to the ultracold energy regime: The 590 MeV, 2.2 mA proton beam coming from the PSI cyclotron (Fig.2a) hits a lead target (Fig.2b) [5], thus creating about 10 neutrons per incident proton. These neutrons are moderated in a heavy water tank at room temperature surrounding the spallation target. A 30 liter solid deuterium crystal ($sD_2$) at 5 K serves to further cool the neutrons and to finally down-scatter the cold neutrons into the UCN regime. Fig.3 shows the container for the $sD_2$ crystal which will be cooled with super-critical helium. Some of the created UCN emanate on top of the crystal into vacuum with an initial energy boost of ~100 neV, corresponding to the material optical potential of $D_2$ [6]. Hydrogen safety made it necessary to have a closed $sD_2$ container which withstands both 3 bar overpressure and 1 bar underpressure. This could be achieved with a 500 μm thin AlMg3 top lid machined from a single piece. UCN can then move 1 m upwards inside a large vertical UCN guide (shown on Fig.5b), and loose the energy from the $D_2$ boost. The vertical guide is a 53 cm inner diameter pure Al tube, vibration-free milled and coated with NiMo on the inside wall. On top, a 2 $m^3$ UCN storage volume coated on the inside with diamond-like carbon (DLC) [7] collects the UCN (see Fig.4). The storage volume will operate at a temperature of roughly 80 K. Once a maximum UCN density is achieved, a large shutter on the bottom of the storage volume is closed and UCN are contained. At this time the proton beam is switched to the other users of the facility. They can then be extracted on demand via shutters on each neutron guide (Fig.6) and then directed via these guides to experiments. In order to do so, the guides have to pass approximately 6 m thick biological shielding surrounding the vacuum tank (Fig.5a) necessary at a high current spallation source. Therefore an excellent UCN transmission is a stringent requirement in order to reach the design density of ~1000 UCN/cm$^3$ at the 2 main experimental ports. UCN guides based on DURAN$^©$ glass tubes with very low surface roughness, inside-coated with NiMo, have been specially developed [8]. In the experimental area South the PSI nEDM experiment has been installed and is awaiting first UCN [3].

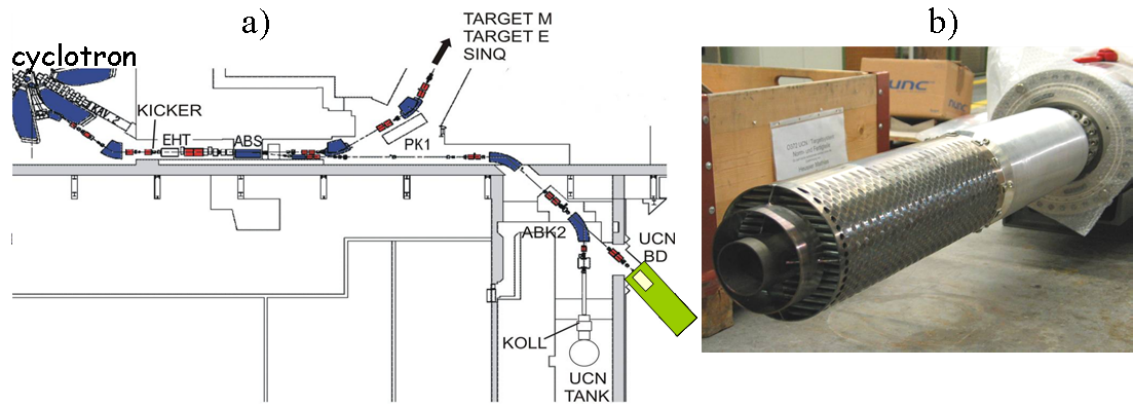

Fig.2: a) Schematics of the UCN relevant part of the PSI proton beamline. Protons are extracted from the 590 MeV cyclotron and after 11 m are kicked [9] onto an extraction beam path towards the UCN spallation target [5], placed in the center of the UCN tank. The UCN beam dump (UCN BD) is used for beam tuning. b) 'Cannelloni' lead spallation target with 756 lead-filled zircaloy rods.

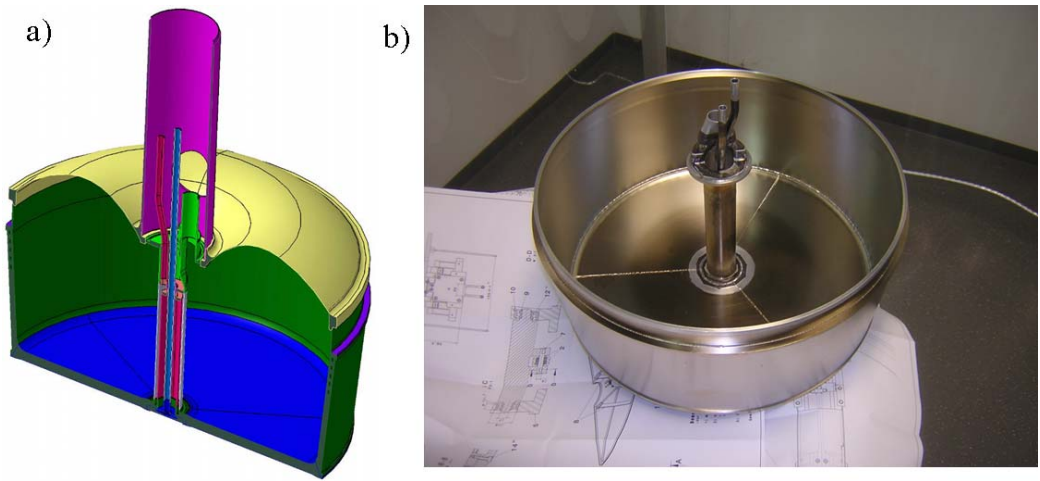

Fig.3: The solid deuterium container made from AlMg3 with a toroidal-shaped 500 μm lid on top. The inside walls are coated with NiMo for better UCN reflection. a) Engineering drawing showing the supply tubes from above for deuterium and He, and the eroded cooling channels inside the wall. b) Inside view of the coated vessel before welding the lid.

A third experimental port extracting UCN from the top of the storage vessel provides a very low energy neutron spectrum with smaller intensity. It is intended for R&D purposes.
A 1 % duty cycle for proton beam operation on the UCN spallation target is foreseen with macro pulses of up to 8 s duration of full beam on target. UCN production will be parasitic with the operation of PSI's pion and muon beam facilities and the SINQ neutron spallation source (99 % duty factor).

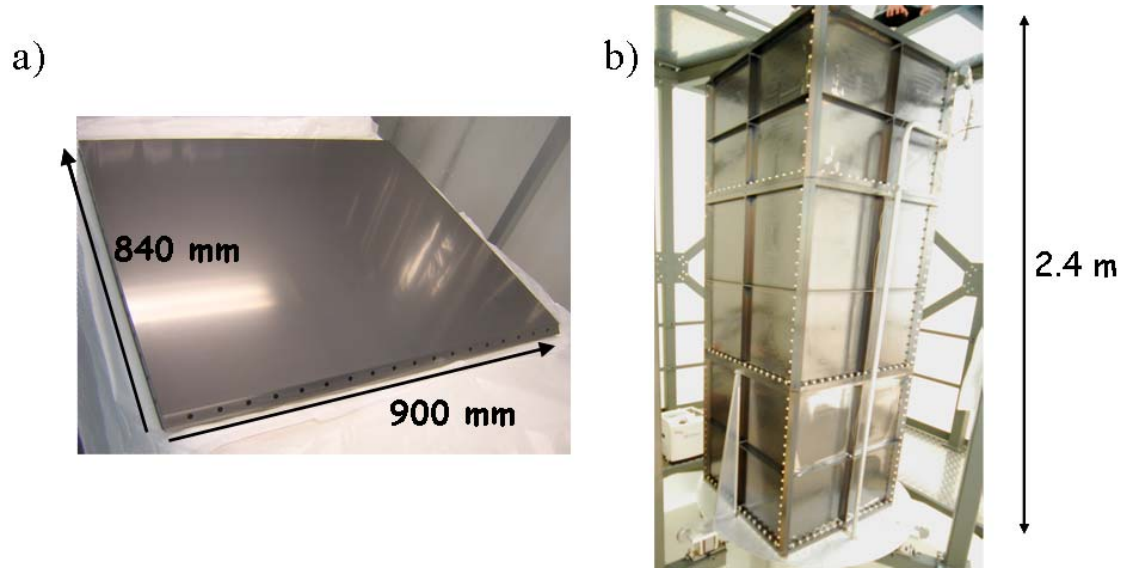

Fig.4: View of the UCN storage volume made form individual Al plates. a) DLC coated wall plate; b) assembled storage volume. The large number of screws serves to minimize cracks.

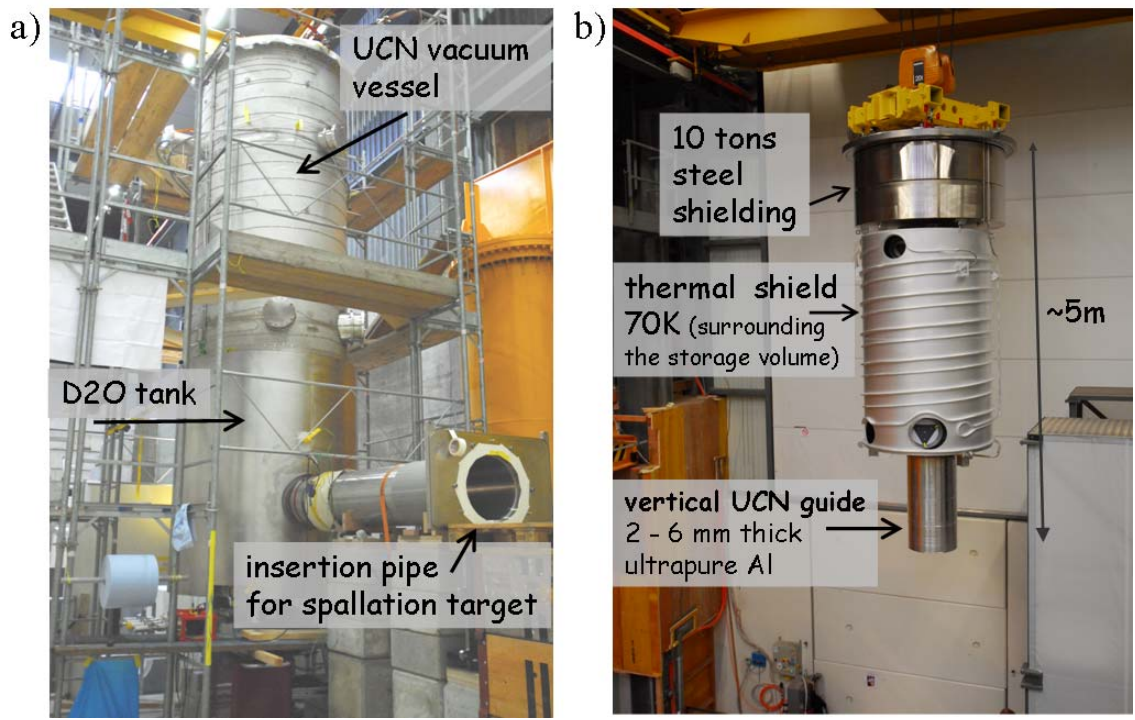

Fig.5: a) UCN vacuum vessel before insertion into the shielding. b) The entire insert of the UCN tank - UCN storage volume surrounded by a thermal shield, vertical UCN guide, top steel shielding - on its way towards final positioning.

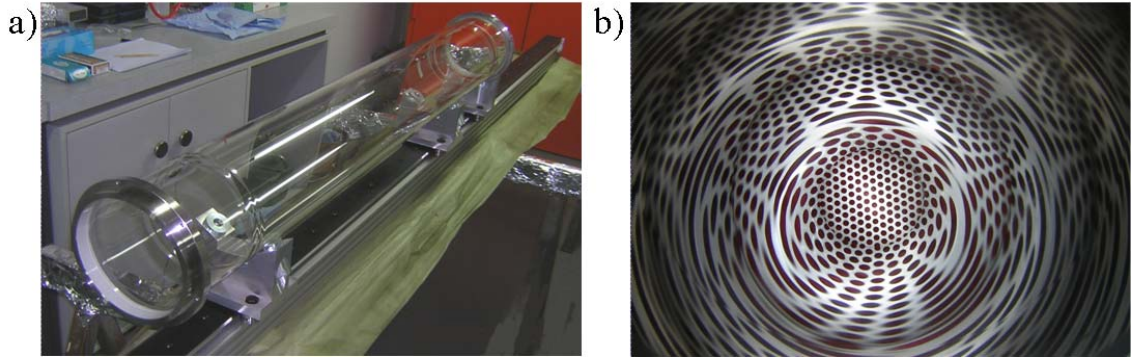

Fig.6: a) Glass UCN guides before coating with NiMo. Steel flanges are glued onto the tube ends for better connectivity [10]. b) Inside view of the polished surface of a steel UCN guide, finally also coated with NiMo.

## 3. Proton beam commissioning

Commissioning of the PSI UCN source started in November 2009 with the heavy water subsystem. This was a prerequisite for the proton beam tests on the spallation target which then followed in December 2009. All UCN beamline elements and beam monitors were tested and performed well. The proton beam was kicked for a maximum of 5 ms onto the UCN spallation target. Proton beam currents were varied between 100 μA and the then allowed full beam current of 2000 μA. A total of 80 beam pulses were directed onto the target.

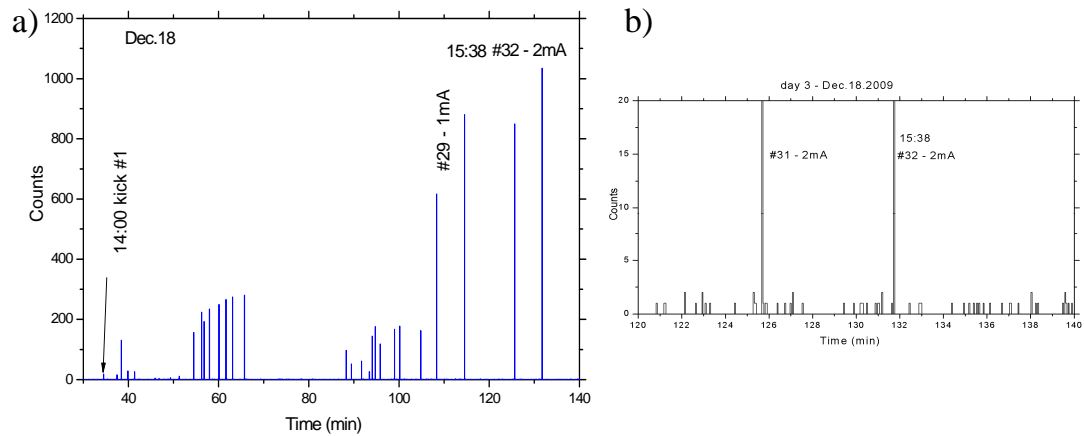

Fig.7: Neutron detection during day 3 of proton beam commissioning. a) All 32 beam kicks of that day on the spallation target were observed in a neutron counter located close to the bend of UCN guide 1-West. Different neutron count rates reflect different beam currents (between 0.3 and 2 mA), beam spreads and beamline magnet settings. b) Zoom of the region between 2 mA kicks. The background rate immediately after and between the beam kicks was measured to be on the same level as before the first beam kick; an important fact for experiments.

During these beam tests we had installed a large CASCADE© neutron detector in the UCN experimental area west. The counter was positioned a the bend of the neutron guide with direct view onto the UCN tank and still important parts of the biological shielding missing. The UCN storage volume was not yet inserted. The counter works via neutron capture in a 200 nm thick $^{10}$B layer deposited on a 20 x 20 cm$^2$ large aluminum foil. The charged particles following neutron capture are then detected in a GEM based gaseous detector operated with an Ar/CO$_2$ mixture. Due to this construction the detector shows a low γ sensitivity. The readout structure divided into 256 sub-pixels allows also for a certain level of background discrimination via pattern recognition. The detector housing was shielded with a 3 mm cadmium layer, which is designed for shielding of cold and thermal neutrons.

Every single proton beam pulse was detected by this counter. Fig.7 shows the observed neutron counts per 3 s time bins on day 3 of the beam commissioning. Some kick numbers for the proton beam are indicated. The last 3 pulses were using the full allowed beam current of 2 mA. At this beam intensity the detector recorded ~1000 counts in one bin coincident with the 5 ms pulse.

Important for any experiment to be performed at the UCN source is the background rate in the detector. Extrapolating the 5 ms, one would expect a 200 kHz rate in a similar detector during a full proton current beam kick of a maximum 8 s length. However, the present detector position was still inside the final shielding. Hence, the expected background will be considerably lower than measured now. Clearly the beam during neutron production will be observed in any experiment. More important however is that the background rate drops immediately after the proton beam kick. We have measured the background before the first ever proton pulse on the UCN target, between the beam kicks and after 3 days of test beam. At this special detector position and in the unfinished shielding configuration we observed a rate of 0.08 ± 0.02 Hz (in a 20 cm$^2$ detection area) at any time but the beam kick time bin. No difference between the various observation periods was found.

The construction of the PSI UCN source is well advanced and scheduled to be finished in fall 2010. Cool down and final commissioning will immediately follow and the first UCN are expected by the end of this year.

**Acknowledgements:** Cordial thanks to the many colleagues contributing to the UCN source project at the Paul Scherrer Institut who are indispensable for the realization of this project. Support of our colleagues at PF2 – ILL and the Mainz TRIGA UCN Source during component testing is acknowledged. PNPI contributed in the early planning of the project.